\documentclass[preprint,superscriptaddress,nofootinbib]{revtex4}

\usepackage{graphicx}
\usepackage{dcolumn}

\usepackage{eucal}
\usepackage[dvips]{epsfig}
\usepackage{amssymb}

\usepackage{amsmath,amsthm}

\newcommand{\bea}{\begin{eqnarray}}
\newcommand{\eea}{\end{eqnarray}}

\newcommand{\be}{\begin{eqnarray}}
\newcommand{\ee}{\end{eqnarray}}

\begin{document}

\title{Rheotaxis facilitates upstream navigation of mammalian sperm cells}

\author{Vasily Kantsler$^{1,2,3}$, J\"orn Dunkel$^{1,4}$, Martyn Blayney$^5$, and Raymond E. Goldstein}
\affiliation{Department of Applied Mathematics and Theoretical Physics, University of Cambridge, 
Wilberforce Road, Cambridge CB3 0WA, UK\\
$^2$Department of Physics, University of Warwick, 
Gibbet Hill Road, Coventry, CV4 7AL, UK\\
$^3$Skolkovo Institute of Science and Technology, Skolkovo, Russia\\
$^4$Department of Mathematics, 
Massachusetts Institute of Technology, 
77 Massachusetts Avenue E17-412, 
Cambridge, MA 02139-4307, USA\\
$^5$Bourn Hall Clinic, Bourn, Cambridge CB23 2TN, UK}

\begin{abstract} 
A major puzzle in biology is how mammalian sperm determine and maintain the correct swimming direction during the 
various phases of the sexual reproduction process.  Whilst chemotaxis is assumed to dominate in the immediate vicinity of 
the ovum, it is unclear which biochemical or physical cues guide  spermatozoa  on their long journey towards the egg cell. 
Currently debated mechanisms range from peristaltic pumping to temperature sensing (thermotaxis) and direct response to 
fluid flow variations (rheotaxis), but little is known quantitatively about their relative importance. Here, we report 
the first quantitative experimental study of mammalian sperm rheotaxis. Using microfluidic devices, we investigate 
systematically  the swimming behavior of human and bull sperm over the whole range of physiologically relevant shear rates and 
viscosities.  Our measurements show that the interplay of fluid shear, steric surface-interactions and chirality of the flagellar 
beat leads to a stable upstream spiraling motion of sperm cells, thus providing a generic and robust rectification mechanism to 
support mammalian fertilisation. To rationalise these findings, we identify a minimal  mathematical model that is capable of 
describing quantitatively the experimental observations. The combined  experimental and theoretical evidence supports the 
hypothesis that the shape and beat patterns of mammalian sperm cells have evolved to optimally exploit rheotaxis for 
long-distance navigation. 
\end{abstract}

\maketitle

The authors declare no competing interests.

%%%%%%%%%%%%%%%%%%%%%%%%%%%%%%%
%\paragraph{General background}
%%%%%%%%%%%%%%%%%%%%%%%%%%%%%%%

\section*{\large{Introduction}}

During their journey from ejaculation to fertilisation, human spermatozoa have to find  
and maintain the right swimming direction over distances that may exceed their head-to-tail length~\mbox{$(\sim100 \mu$m)} 
by a thousandfold.  On their path to the egg cell, mammalian sperm encounter varied physiological environments, are exposed to 
a variety of chemical gradients and must overcome counter-flows. Whilst chemotactic sensing~\cite{2008Kaupp_AnnRev} is assumed to 
provide important guidance in the immediate vicinity of the ovum~\cite{2003Spehr_Science}, it is not known which 
biochemical~\cite{2012Brenker_EMBO} or physical mechanisms~\cite{1984Winet} keep the sperm cells on track as they pass through the 
rugged landscapes of cervix, uterus and oviduct~ \cite{2005Katz,2006Suarez,2006Eisenbach_NatRev}. The complexity of the mammalian 
reproduction process and not least  the lack of quantitative data make it very difficult to
assess the relative importance of the various proposed long-distance navigation mechanisms~\cite{2006Fauci_AnnRev,2006Eisenbach_NatRev}, 
ranging from cervix contractions~\cite{2006Suarez,2006Fauci_AnnRev} and chemotaxis~\cite{2008Kaupp_AnnRev} to 
thermotaxis~\cite{2003Bahat_NatMed} and rheotaxis~\cite{2013Miki_CurrBiol}.  Aiming to understand not only qualitatively but also 
quantitatively how  fluid-mechanical effects may help steer mammalian spermatozoa over large distances, we report 
here a combined 
experimental and theoretical study of sperm swimming in microfluidic channels, probing a wide range of physiologically relevant 
conditions of shear and viscosity. For both human and bull spermatozoa, we find that their physical response to shear flow, combined 
with an effective shape-regulated surface attraction~\cite{2013Kantsler_PNAS} and head-tail counter-precession, favors an upstream 
spiraling motion along channel walls. The  robustness of this fluid-mechanical rectification mechanism suggests that it is likely to 
play a key role in the long-distance navigation of mammalian sperm cells. Thus, the detailed analysis reported below not only yields 
new quantitative insights into the role of biophysical processes during mammalian reproduction but could also lead to new diagnostic 
tools and improved artificial insemination techniques~\cite{2010IUF_review}.

%%%%%%%%%%%%%%%%%%%%%%%%%%%
%\paragraph{More specific background}
%%%%%%%%%%%%%%%%%%%%%%%%%%%
Recent experiments on red abalone~\cite{2007Riffell_JEB,2011Zimmer_PNAS}, a large marine snail that fertilises externally, 
showed that weak fluid flows can be beneficial to the reproduction of these organisms, suggesting that shear flows could have 
acted as a selective pressure in gamete evolution. In higher organisms, which typically fertilise internally, sperm transport 
is much more complex and the importance of shear flows relative to chemotaxis, peristaltic pumping or thermotaxis still poses 
an open problem as it is difficult to perform well-controlled~\textit{in vivo} studies. The complex uterine and oviduct 
topography~\cite{2006Suarez} and large traveling distances render it unlikely that local chemotactic gradients steer sperm 
cells during the initial and intermediate stages of the sexual reproduction process.  Experimental evidence~\cite{1996Kunz_HumanReprod} 
suggests that rapid sperm transport right after insemination is supported by peristaltic pumping driven by muscular 
contractions of the uterus, but it is not known how sperm navigate in the oviduct.  Thermotaxis, the directed 
response of sperm to local temperature differences, was proposed as a possible long-range rectification mechanism of 
sperm swimming in rabbits~\cite{2003Bahat_NatMed}, but recently questioned~\cite{2013Miki_CurrBiol} as it is likely to 
be inhibited by convective currents that form in the presence of temperature gradients.  On the other hand, it has 
long been known that,  similar to bacteria~\cite{2012Stocker_PNAS} and algae~\cite{2013Chlamy_PRE},  mammalian 
sperm~\cite{1905Adolphi,1970Roberts} are capable of performing rheotaxis, by aligning against a surrounding 
flow~\cite{2012Stocker_PNAS,2013Miki_CurrBiol}, but this effect has yet to be systematically quantified in 
experiments~\cite{2006Suarez,2006Fauci_AnnRev}.  Specifically, it is not known at present how sperm 
cells respond to variations in shear rate and viscosity, and how long they need to adapt to temporal changes 
in  the flow direction. Answering these questions is essential for understanding which physical effects may be important  
at different stages of the mammalian fertilisation process.

%%%%%%%%%%%%%%%%%%%%%%%%%%%%%
%\paragraph{Main results of this work}
%%%%%%%%%%%%%%%%%%%%%%%%%%%
To quantify the swimming strategies of sperm cells under well-controlled flow conditions, we performed a series of 
microfluidic experiments in cylindrical and planar channels (Fig.~1), varying systematically shear rates $\dot{\gamma}$ and viscosities $\mu$ through the physiologically and rheotactically relevant regime, up to $\mu=20$mPa$\cdot$s which is roughly $10\times$ the viscosity of natural seminal fluid~\cite{2005Owen}. These measurements revealed the interesting result that both human and bull 
spermatozoa do not simply align against the flow, but instead swim upstream on spiral-shaped trajectories along the 
walls of a cylindrical channel (Fig.~1A and Movie~1). The previously unrecognised transversal velocity component can be 
attributed to the chirality of the flagellar beat. The resulting helical swimming patterns enable the spermatozoa to explore 
collectively the full surface of a cylindrical channel, suggesting that rheotaxis can help sperm to navigate their way 
through the oviduct and find the egg cell~\cite{2013Miki_CurrBiol}.  Using high-speed imaging, we also determined the dynamical 
response of human and bull spermatozoa to flow reversal at different viscosities, which is essential for understanding how active 
swimming, rheotaxis and uterine peristalsis can combine to facilitate optimal sperm transport. To rationalise the experimental 
observations, we identify below a simple mathematical model that reproduces the main results of our measurements.

%%%%%%%%%%%%%%%%%%%%%%%%%%%%%%%%%%%
\section*{\large{Results}}
%%%%%%%%%%%%%%%%%%%%%%%%%%%%%%%%%%%
\subsection*{Shear \& viscosity dependence}
%%%%%%%%%%%%%%%%%%%%%%%%%%%%%%%%%%%

In the experiments, samples of human and bull spermatozoa were injected into microfluidic channels of spherical or rectangular 
cross-section (Methods). The cells were then exposed  to  well-defined Poiseuille shear flows, corresponding to parabolic 
flow profiles (Fig.~1B).  Even in the absence of flow,  sperm cells tend to accummulate at surfaces~\cite{1963Rothschild_Nature,2012Denissenko_PNAS} due to a combination of steric repulsion~\cite{2013Kantsler_PNAS} and  hydrodynamic forces~\cite{1995Fauci_BMB,2010Elgeti_BiophysJ,2011Gaffney_AnnRev,2010Friedrich_JEB,2012Blake_EPJE}. 
This can be explained by the fact that,  in essence,  the flagellar beat traces out a cone which, upon collision, 
aligns with a solid surface, so that the sperm's propulsion vector points into the boundary and the cells become 
effectively trapped at the surface~\cite{2013Kantsler_PNAS}. In the presence of a Poiseuille shear flow, cells close to the channel boundaries 
experience an approximately linear vertical flow profile,  whose slope is given by the shear rate~$\dot\gamma$ (Fig.~1B).  
To quantify the effects of shear rate and viscosity on sperm swimming, we tracked a large number of individual cells 
(typically $N>10,000$) in planar microfluidic channels (Fig.~1C) at different shear rates $\dot\gamma$, ranging from 0.2 s$^{-1}$ to 9 s$^{-1}$, and different dynamic viscosities $\mu$, ranging from 1mPa$\cdot$s (that of water) to  20mPa$\cdot$s (see Methods). The cell tracks were then used to reconstruct the  velocities of sperm swimming close to the boundary.  Mean values and histograms of the upstream and transverse velocity components from those measurements are summarised in Figs.~1D,E.  Since sperm motility depends on viscosity and may vary among different samples, it is advisable to normalise sperm velocities $\boldsymbol{v}=(v_x, v_y)$, that have been measured at different values of $\dot\gamma$ and $\mu$, by the mean sample speed $v_{0\mu}=\langle|\boldsymbol v| \rangle_{\mu} $ at zero shear $\dot\gamma=0$, and  also to rescale the flow velocity accordingly. Figure~1D shows the thus-normalised mean upstream and mean transverse swimming velocities $\langle v_{y,x} \rangle_\mu/v_{0\mu}$ for bull and human spermatozoa as a function of the dimensionless rescaled shear flow speed $u_{20}=\dot\gamma (2A/v_{0\mu})$, where $A=10\mu$m is the approximate amplitude of a typical flagellar beat (i.e. the maximum distance  of the flagellar tip from the surface during a beat is $2A$).  The results for the upstream velocity reveal that both human and bull sperm exhibit optimal upstream swimming at  rescaled flow speeds $u_{20}\sim 1$, implying that there is an optimal shear regime for the rectification of sperm swimming. Remarkably, however, we also find that, at low viscosities ($\mu\ll$10mPa$\cdot$s), human spermatozoa exhibit a substantial  shear-induced transverse velocity component that becomes suppressed at very high viscosities. By contrast, for bull spermatozoa, the mean transverse component is generally weaker and less sensitive to viscosity variations. These statements are also corroborated by the corresponding velocity histograms in Fig.~1E. 
\par
Qualitatively, the above observations for stationary shear flows can be explained as follows.  Once a sperm cell has 
become trapped at a surface, its tail explores, on average, regions of higher flow velocity than the head, resulting in 
a net torque that turns the head against the flow (Fig.~1B). This shear-induced rectification is counter-acted by variability 
in the cells swimming direction. If the shear velocity is too low the orientational \lq noise\rq, which is caused by a combination of intrinsic fluctuations in the cells' swimming apparatus, thermal fluctuations and elastohydrodynamic effects,  inhibits upstream 
swimming, whereas if the  shear velocity  becomes too large the sperm will simply be advected downstream by the flow, 
implying that there exists an optimal intermediate  shear rate for upstream swimming. Interestingly, we find that the maximum 
of the upstream velocity decreases more strongly with viscosity for bull sperm than for human sperm (Fig.~1D).  This could be due to differences in cell morphology, as previous numerical studies~\cite{2011Gaffney_BiophysJ_Comment} for bacterial cells suggest that differences in head shape can substantially alter swimming behavior. Bull sperm have a flatter head than human sperm, which likely suppresses the rotational motion of the cell at high viscosities thus leading to an effectively smaller vertical beat amplitude~$A$. This could explain why, at high values of $\mu$, the tail beat of bull sperm becomes essentially two-dimensional and constricted to the vicinity of the surface, so that alignment against the flow becomes less efficient.

%%%%%%%%%%%%%%%
To understand the unexpectedly strong transverse velocity component of human sperm 
at $\mu\lesssim 5$mPa$\cdot$s, as typical of the seminal fluid~\cite{2005Owen}, it is important to recall 
that sperm of invertebrae and mammals are known to exhibit different chiral beat patterns depending on environmental 
conditions~\cite{1982Gibbons,1993Ishijima_BiophysJ,2001Woolley_JEB,2009Smith}, and that shear flows are capable of separating particles 
along the transverse direction according to their chirality~\cite{2009Stocker_PRL,2012Talkner_NJP}. Human sperm exhibit a 
strongly helical beat component at low-to-moderate values of $\mu$ (Movie 2), but this chirality becomes suppressed at high viscosities~(Movie~3) resulting in more planar wave forms~\cite{2009Smith}.
For comparison, the beat of a bull sperm flagellum is more similar to a rigidly rotating 
planar wave even at low viscosities~(Movie 4), thus exhibiting a weaker chirality and leading to smaller transverse velocities~(Fig.~1D). Since the flagellar beating pattern can be controlled not only by viscosity but also by changes in Calcium 
concentration~\cite{1993Ishijima_BiophysJ}, higher organisms appear to possess several means for tuning transverse and 
upstream swimming of sperm.

%%%%%%%%%%%%%%%%%%%%%%%%%%%%%%%%%%%
\subsection*{Dynamical response}
%%%%%%%%%%%%%%%%%%%%%%%%%%%%%%%%%%%

In addition to typically outward directed mucus flow in the oviduct, sperm cells are also exposed to temporally varying flows 
driven by uterine contractions~\cite{2006Suarez,2006Fauci_AnnRev}. To probe the dynamical response of sperm to changes in the 
flow direction, we performed additional experiments where we tracked the motion of bull and human spermatozoa after a sudden 
flow reversal  at two different viscosities (Fig.~2 and Movies~5, 6). In those experiment, sperm were first given time to 
align against a stationary shear flow, then the flow direction was reversed, $u_y\to-u_y$, with a switching time $< 1$s. Upon 
flow reversal, a sperm cell typically performs a U-turn (Fig.~2A,B). The characteristic radius of curvature of the trajectory and 
the typical turning time $\tau$ were found to increase strongly with viscosity. At low viscosity, $\mu\sim 1$mPa$\cdot$s, sperm 
realign rapidly against the new flow direction with a typical response time of $\tau\sim 5$s to 10s, and the curvature radius 
is of the order of one or two sperm lengths $\ell\sim 60\mu$m (Movie~5). By contrast, at a larger viscosity of 
$\mu\sim 12$mPa$\cdot$s, which is roughly $4\times$ higher than the natural viscosity of the ejaculate,  both curvature 
radius and response time increase by approximately a factor of $5$ (Movie~6). Interestingly, these response times are of 
the order of typical cervical contractions~\cite{1996Kunz_HumanReprod}, suggesting a possible fine-tuning between muscular 
activity of the uterus and turning behavior of sperm cells. In particular, immediately after the flow reversal, sperm 
orientation and flow direction point for a short period of time in approximately the same direction, leading to a 
momentarily increased transport velocity (see velocity peaks in Fig.~2C). Thus,  by switching flow directions back 
and forth at an optimal rate, the transport efficiency of an initially rectified sperm population can be enhanced.

%%%%%%%%%%%%%%%%%%%%%%%%%%%%%%%%%%%
\subsection*{Minimal model}
%%%%%%%%%%%%%%%%%%%%%%%%%%%%%%%%%%%

To test whether our understanding of the experimental observations is correct and to provide a basis for future theoretical 
studies, we used resistive force theory to infer a minimal mathematical model that incorporates  the main 
physical mechanisms discussed above (details are provided in the Supplementary File 1). The model assumes that the effectively 
two-dimensional motion of a sperm cell,
that swims close to a surface in the presence of a shear flow can be described in terms of its  position vector 
$\boldsymbol R(t)=(X(t),Y(t))$ and its orientation unit vector $\boldsymbol N(t)=(N_x(t),N_y(t))$. Focussing on an 
effective description of the main physical effects and assuming that the flow is in $y$-direction (Fig.~1B), the equations 
of motions for $\boldsymbol R$ and $\boldsymbol N$ read 
\begin{eqnarray}
\label{eq1}
\dot{\boldsymbol R} 
&=& V\boldsymbol N + \sigma\overline{U}{\boldsymbol e}_y,
\\
\dot{\boldsymbol N}
&=& 
\sigma\dot{\gamma}\alpha
\begin{pmatrix}
N_xN_y\\
N_y^2-1
\end{pmatrix}
+
\sigma\dot{\gamma}\chi  \beta
\begin{pmatrix}
N_x^2- 1\\
N_x N_y 
\end{pmatrix}
+(2D)^{1/2}(\boldsymbol I-\boldsymbol N\boldsymbol N)\cdot\boldsymbol\xi(t).
\label{eq2}
\end{eqnarray}
Equation~(1) states that the net in-plane velocity $\dot{\boldsymbol R}(t)$ of a cell arises from two main contributions:  
self-swimming  at typical speed $V$ in the direction of the cell orientation $\boldsymbol N$,  and advection 
by the flow, where $\sigma =\pm 1$  defines the flow direction and $\overline{U}>0$ the mean flow speed experienced by the 
cell. As explained in detail in the Supplementary File 1, the nonlinear structure of Eq.~(2) ensures that  the length of the orientation 
vector $\boldsymbol N$ remains conserved,  assuming that the change in orientation, $\dot{\boldsymbol N}(t)$, is caused by 
three effects:  shear-induced alignment against the flow with rate $\dot{\gamma}\alpha$ where $\alpha>0$ is numerical 
factor that encodes geometry of the flagellar beat, shear-and-chirality-induced turning at rate $\dot{\gamma}\beta$ 
with $\chi\in\{-1,0,+1\}$ and $\beta>0$ encoding chirality and shape of the flagellar beat, and variability~\cite{2012Su_PNAS} in the swimming direction, modeled as a Stratonovich-type two-dimensional Gaussian white noise  $\boldsymbol \xi$ with amplitude $D$~\cite{2006Lubensky}.  Equations~(1) and~(2)  were obtained by approximating the flagellum by a rigid conical helix, with the polar geometry of the enveloping cone dictating the mathematical structure of the deterministic turning terms  (see Supplementary File 1 for details of the calculation). The simplifying assumptions underlying Eqs.~(1) and~(2) imply that this minimal model does not accurately capture the dynamics of individual cells at zero shear, as the deterministic terms in Eq.~(2) neglect the intrinsic curvature of cell trajectories. However, when analyzing the in-plane curvature for a large number of human sperm trajectories ($>$100,000 sample points from more than 1200 cells) at zero shear, we found a broad distribution of curvatures with a small positive mean curvature of $(5.6\pm 1.3)\cdot 10^{-4}\,\mu$m$^{-1}$ at low viscosity (1mPa$\cdot$s) and  a small negative mean curvature  $(-1.9\pm 0.1)\cdot 10^{-3}\,\mu$m$^{-1}$ at high viscosity (12mPa$\cdot$s), where the different signs are consistent with the observed change in the transverse velocity for human sperm at high viscosity (Fig.~1D). To account at least partially for these curvature variations, we include in Eq.~(2) the Gaussian white noise term. Compared with more accurate models that resolve the details of the flagellar dynamics~\cite{2010Elgeti_BiophysJ,2011Gaffney_AnnRev}, Eqs.~(\ref{eq1}) and~(\ref{eq2}) provide a strongly reduced description which, however, turns out be sufficient for rationalising our experimental observations (Fig.~3). Values for $V$ and $\overline U$ can be directly estimated from experiments, and sign conventions in Eq.~(2) 
have been chosen such that $\chi =+1$ for human sperm at low viscosity (for weakly chiral bull sperm one can use $\chi= 0$ in 
a first approximation). The  model parameters $(\alpha,\, \beta,\, D)$ can be inferred from the experimental data (Supplementary File 1). By 
performing systematic parameter scans,  we found that values $\alpha\in [0.2, 0.4]$, $\beta\in [0.05, 0.1]$  and 
$D\in [0.2,0.3]\,$rad$^2$/s yield good quantitative agreement with the experimental results for both stationary flow (Fig.~3A) 
and flow reversal (Fig.~3B), suggesting that the coupling between shear flow and beat chirality dominates the transverse velocity dynamics. 
We may therefore conclude that, despite some strong simplifications, the effects included in the model capture indeed the main physical 
mechanisms relevant for understanding sperm motion in shear flow near a surface.

%%%%%%%%%%%%%%%%%%%%%%%%%%%%%%%%%%%
\section*{\large{Discussion}}
%%%%%%%%%%%%%%%%%%%%%%%%%%%%%%%%%%%

In conclusion, we have reported detailed quantitative measurements of sperm motion in shear flow. 
Our experimental results show that upstream swimming of mammalian sperm due to rheotaxis is more 
complex than previously thought. Human sperm cells were found to exhibit a significant 
transverse velocity component that could be of relevance in the fertilisation process, as the ensuing 
spiraling motion enables spermatozoa to explore collectively a larger surface area of the oviducts, thereby 
increasing the probability of locating egg cells. Our theoretical analysis implies that the transverse velocity 
component arises from a preferred handedness in the flagella beat in the presence of shear flow, in contrast to 
recent findings for male microgametes of the malaria parasite~\textit{Plasmodium berghei}~\cite{2013Wilson_PNAS}. 
Due to the large sample size, our data provide substantial statistical evidence for the hypothesis that mammalian 
sperm have evolved to achieve optimal upstream swimming near surfaces, possibly exploiting the enhanced fluid production in the female reproductive system during the fertile phase~\cite{2000Eschenbach} and after intercourse~\cite{2013Miki_CurrBiol}. 
The improved 
quantitative knowledge derived from this data may help to design more efficient artificial insemination strategies, 
for example, by optimising the viscosity and chemical composition of fertilisation media and adjusting injection 
techniques to maximise upstream swimming of sperm cells. Combined with recent measurements~\cite{2013Kantsler_PNAS}, 
which clarified the importance of flagella-mediated contact interactions for the accumulation of sperm cells at surfaces, 
the results presented here yield a cohesive picture of the mechanistic and fluid-mechanical~\cite{2010Friedrich_JEB} 
aspects of long-distance sperm navigation. Future work should focus on merging these insights with quantitative studies of 
chemotaxis~\cite{2003Spehr_Science,2008Kaupp_AnnRev,2011Zimmer_PNAS,2012Brenker_EMBO} to obtain a differentiated understanding 
of the interplay between physical and chemical factors during various stages of the mammalian reproduction process.

%%%%%%%%%%%%%%%%%%%%%%%%%%%%%%%%%%%%%%%
\section*{\large{Methods}}
\label{methods}
%%%%%%%%%%%%%%%%%%%%%%%%%%%%%%%%%%%%%%%

\subsection*{Sperm sample preparation}
Cryogenically frozen bull spermatozoa were purchased from Genus Breeding. 
For each experiment, a bull sperm sample of 250 $\mu$L was thawed in a water bath at 37$^\circ$C for 15 sec. Human samples 
from healthy undisclosed normozoospermic donors were obtained from Bourn Hall Clinic. Donors provided informed consent 
in accordance with the regulations of The University of Cambridge Human Biology Research Ethics Committee. For each 
experiment, bull and human samples were washed three times by centrifugation at 500 rcf for 5 min with the appropriate 
medium. The medium for bull spermatozoa contained 72 mM KCl, 160 mM sucrose, 2 mM Na-pyruvate, and 2 mM Na-phosphate buffer 
at pH 7.4~\cite{2001Woolley_JEB}. Human sperm medium was based on a standard Earle's Balanced Salt Solution, containing 66.4 
mM NaCl, 5.4 mM KCl, 1.6 mM CaCl$_2$, 0.8 mM MgSO$_4$, N$_2$H$_2$PO$_4$ 1 mM, NaHCO$_3$ 26 mM, D-Glucose 5.5 mM supplemented 
with 2.5 mM Na pyruvate and 19 mM Na-lactate pH adjusted to 7.2 by bubbling the medium with CO$_2$.  Viscosity of the medium 
was modified by adding methylcellulose (M0512; Sigma-Aldrich; approximate molecular weight 88,000) at concentrations 
0\%, 0.2\%, 0.4\%, 0.5\% w/v. The absence of circular trajectories at zero-shear implies that the sperm are 
capacitated~\cite{2013Miki_CurrBiol}.

\subsection*{Microfluidics}
Microfluidic channels were manufactured using standard soft-lithography techniques. The master mould was produced 
from SU8 2075 (MicroChem Corp.) spun to a 340 microns thickness layer and exposed to UV light through a high resolution 
mask to obtain the desired structures. The microfluidic chip containing the channels cast from PDMS (Sylgard 184, 
Dow Corning) and bonded to covered glass. The channel has rectangular cross-section of 0.34$\times$3 mm. We treated 
PDMS surfaces of the channels prior the experiment with 10\% (w/v) Polyethylene glycol (m.w. 8000, Sigma) solution 
in water for 30 min to avoid adhesion of sperm cells to the walls. Sperm suspension were introduced through inlets 
with a micro-syringe pump (Harvard Instruments) at  controlled flow rates of 0.1 to 40 $\mu$L/min. The concentration 
of the sperm cells in the experiments was kept below 1\% volume fraction. 

\subsection*{Microscopy}
To identify the swimming characteristics of individual sperm cells, the trajectories were reconstructed by applying a 
custom-made particle-tracking-velocimetry (PTV) algorithm to image data taken with a Zeiss Axio Observer inverted 
microscope (20x or 10x objective, 25 fps). The flagella dynamics was captured with a Fastcam SA-3 Photron camera 
(125 fps, 40x/NA 0.6 objective). Calibration of the velocity profile in the channel was performed by measuring 
trajectories of fluorescent beads for different distances from the coverslip via PTV. The measured velocity profile 
is found identical to with the calculated values from solving the Stokes equations for the given geometry. Values of the 
shear rate $\dot\gamma$ in the different experiments were reconstructed from the flow velocity at distance 20 $\mu$m 
from the wall (see below).

\subsection*{Additional experimental information} 

Effects of viscosity variation and shear-rate variation  were studied in experiments that were performed in a rectangular channel with a cross-section 0.34$\times$3mm$^2$, by observing sperm motion at lower and upper channel walls. The field of view (normally $800\times800\mu$m$^2$) was chosen at the middle of the channel (in $x$-direction), where the in-plane velocity gradient is negligible due to the high aspect ratio of the channel (see Fig.~\ref{fig:flow}B). The $v_y$-velocity profiles, measured along the $z$-coordinate, was found to be in perfect agreement with the theoretically expected parabolic flow profile for this geometry (see Fig.~\ref{fig:flow}B). The shear rate $\dot\gamma$  at a given flow rate was determined from the flow velocity at distance 20 $\mu$m from the wall. The depth of field of the objective was $< 5\mu$m to ensure that we only observed cells that swam close to the surface. Trajectories of individual sperm cells were analysed in MATLAB. The sample size in a single experiment exceeds 100,000 velocity vectors, each measurement for a given viscosity and a shear rate was repeated a few times with different sperm samples.  Available on request are $8$ supplemental data tables that summarise the statistical information for each experiment.

%%%%%%%%%%%%%%%%%%%%%%%%%%%%%%%%%%%%%%%%
%%%%%%%%%%%%%%%%%%%%%%%%%%%%%%%%%%%%%%%%
%%%%%%%%%%%%%%%%%%%%%%%%%%%%%%%%%%%%%%%%
%\bibliography{Sperm}

%%%%%%%%%%%%%%%%%%%%%%%%%%%%%%%%%%%%%%%%
%%%%%%%%%%%%%%%%%%%%%%%%%%%%%%%%%%%%%%%%
%%%%%%%%%%%%%%%%%%%%%%%%%%%%%%%%%%%%%%%%

%%%%%%%%%%%%%%%%
\vspace{0.5cm}
\textbf{Additional information}\\
\noindent
Correspondence and request for materials should be addressed to R.~E.~G.

%%%%%%%%%%%%%%%%
\section*{Movie Information}

\paragraph*{Movie 1}
Human sperm cell swimming on a spiral trajectory (green) against a shear flow in a cylindrical channel (fluid viscosity 3mPa$\cdot$s; channel diameter 300 $\mu$m; channel boundaries marked in red). Scale bar $100\,\mu$m.

\paragraph*{Movie 2}%to be made
Human sperm cells swimming in a low-viscosity fluid (3mPa$\cdot$s) near the wall of a planar channel. The movie shows that, at low viscosity, the flagellar 
beat of a human sperm cell typically exhibits a considerable chiral component. 
This follows from the fact that the flagellum never appears as a straight line 
(in contrast to bull sperms at same viscosity, compare Movie 4). Scale bar 20$\,\mu$m. 

\paragraph*{Movie 3} 
Human sperm cells swimming in a high-viscosity fluid (20mPa$\cdot$s) near the wall of  a planar channel. The movie shows that, at very high viscosity, the chiral 
beat component becomes considerably weaker for there now exist instances where the flagellum appears as an almost straight line, indicating that the beat pattern approaches the shape of a planar rotating wave. Scale bar 20$\,\mu$m.

\paragraph*{Movie 4} %change labels and scalar bar info
Bull sperm cells swimming in a low-viscosity fluid (3mPa$\cdot$s) near the wall of  a planar channel.  The movie shows that, even at low viscosity, the flagellar beat of bull sperm is approximately planar. This follows from the fact that at certain instances the flagellum appears as a line (in contrast to human sperms at same viscosity, compare Movie~2).
Scale bar 20$\,\mu$m.

\paragraph*{Movie 5}
Reorientation of a human sperm cell swimming in a low-viscosity fluid (1mPa$\cdot$s) in a planar channel, after a sudden reversal of the flow direction at time $t=0$.

\paragraph*{Movie 6}
Reorientation of two human sperm cells, swimming in a high-viscosity fluid in a planar channel, after a sudden reversal of the flow direction at time $t=0$.

%%%%%%%%%%%%%%%%
\section*{Additional files}

\paragraph*{Supplementary File 1}  
This file contains a detailed mathematical derivation of the minimal model in Eqs.~(1) and~(2) of the main text, a description of the parameter estimation procedure and a brief summary of numerical methods.

%%%%%%%%%%%%%%%%%%%%%%%%%%%%%%%%%%%%%%%
%%%%%%%%%%%%%%%%%%%%%%%%%%%%%%%%%%%%%%%
\begin{figure*}
\centering
\includegraphics[width=10.5cm]{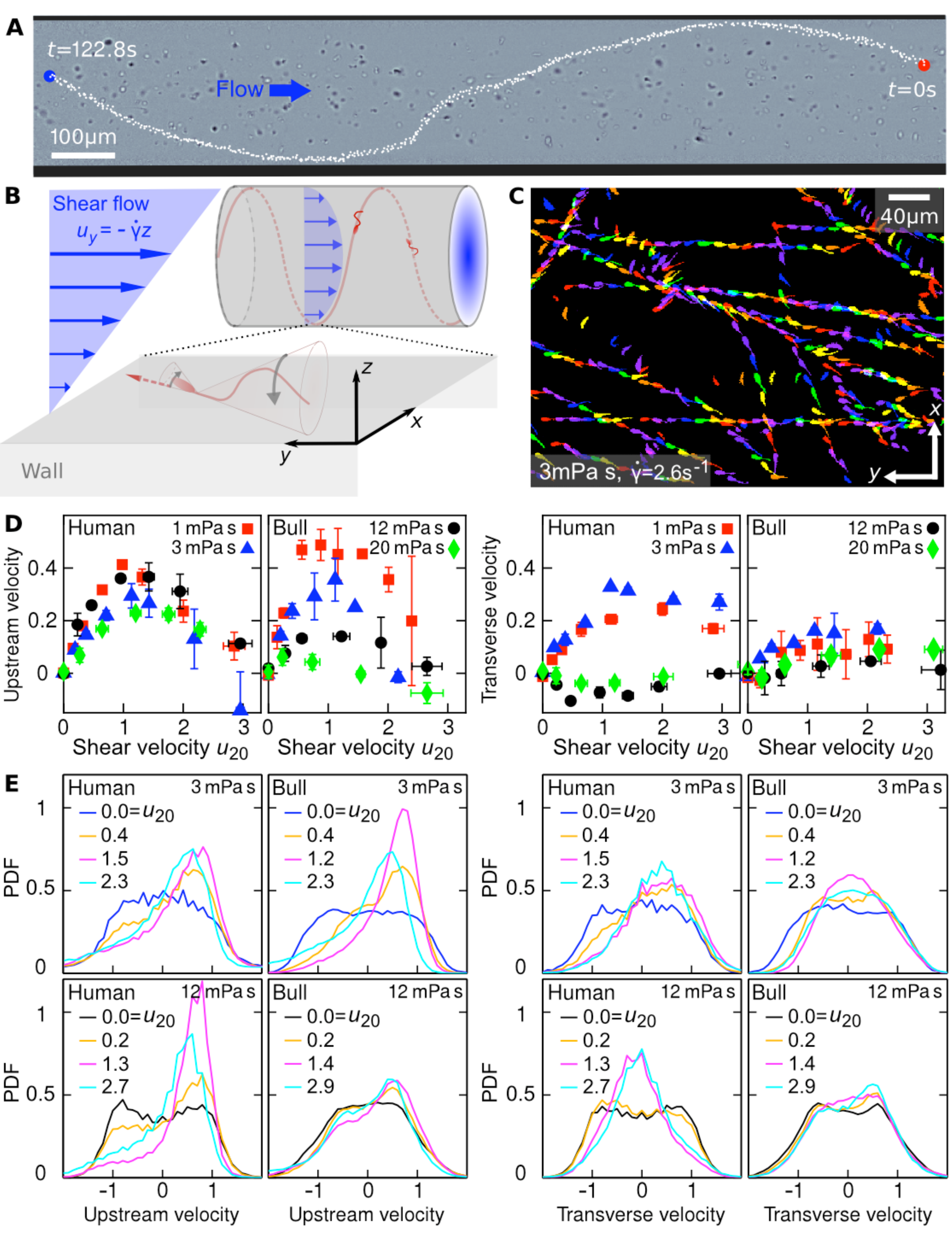}
\caption{
Sperm swim on upstream spirals against shear flow.
(A)~Background-subtracted micrograph showing the track of a bull sperm in a cylindrical channel (viscosity $\mu=3\,$mPa$\cdot$s 
shear rate $\dot{\gamma}=$2.1\,s$^{-1}$), channel boundary false-colored with black, see Movie~1 for raw data. (B)~Schematic 
representation not drawn to scale. The conical envelope of the flagellar beat holds the sperm close to the surface~\cite{2013Kantsler_PNAS}. 
The vertical flow gradient exerts a 
torque that turns the sperm against the flow, but is counteracted by a 
torque from the chirality of the flagellar wave, resulting in a mean diagonal upstream   
motion. (C)~Tracks of bull sperm near a flat channel surface.
(D) Upstream and transverse mean velocities $\langle v_{y,x}\rangle$ versus shear flow speed~$u_{20}$ at $20\,\mu$m from 
the surface for different viscosities. All velocities are normalised by the sample mean speed $v_{0\mu}$ at~$\dot\gamma=0$.  
For human sperm, in order of increasing viscosity $v_{0\mu}=53.5\pm3.0,\;46.8\pm 3.7,\;36.8\pm 3.3,\;  29.7\pm 3.9\,\mu$ms$^{-1}$, 
and for bull sperm   $v_{0\mu}= 70.4 \pm 11.8,\, 45.6\pm 4.7,\;32.4\pm 4.8,\, 29.6\pm 4.1\,\mu$ms$^{-1}$, where uncertainties are 
standard deviations of mean values from different experiments. Each data point is an average over $>1000$ sperms. 
(E) Histograms for selected points in (D).
}
\end{figure*}
%%%%%%%%%%%%%%%%%%%%%%%%%%%%%%%%%%%%%%%
%%%%%%%%%%%%%%%%%%%%%%%%%%%%%%%%%%%%%%%
\newpage
%%%%%%%%%%%%%%%%%%%%%%%%%%%%%%%%%%%%%%%
%%%%%%%%%%%%%%%%%%%%%%%%%%%%%%%%%%%%%%%
\begin{figure*}[t]
\centering
\includegraphics[width=12cm]{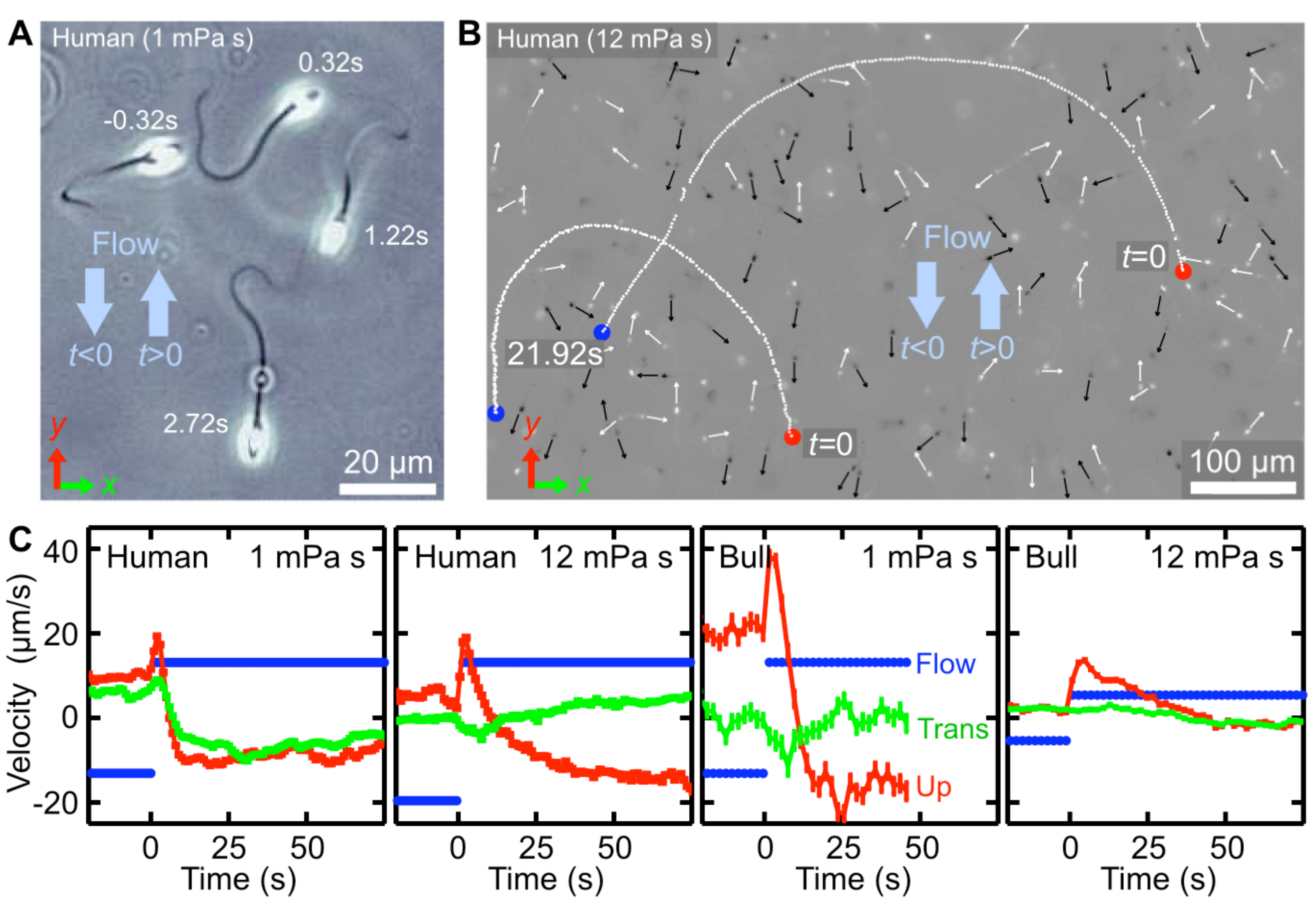}
\caption{
Temporal response of sperm cells to a reversal of the flow direction depends sensitively on viscosity.  (A) 
At low viscosity, sperm perform sharp U-turns, see also video 2.  (B) At high viscosity, the typical radius of the 
U-turns increases substantially (Movie 3). 
 White/black arrows show orientations of  several cells before/after turning.
(C)~Flow velocity at distance $5\;\mu$m from the channel surface (blue, \lq Flow\rq), mean upstream velocity $\langle v_y\rangle$ 
(red, \lq Up\rq) and  mean transverse velocity $\langle v_x\rangle$ (green, \lq Trans\rq) as function of time. 
The typical response time of sperm cells after flow reversal increases with viscosity. Peaks reflect a short period when mean 
swimming direction and flow direction are aligned. The time series for human sperm also signal a suppression of the beat 
chirality at high viscosity, consistent  with~Fig.~1D.}
\end{figure*}
%%%%%%%%%%%%%%%%%%%%%%%%%%%%%%%%%%%%%%%
%%%%%%%%%%%%%%%%%%%%%%%%%%%%%%%%%%%%%%%
\newpage

%%%%%%%%%%%%%%%%%%%%%%%%%%%%%%%%%%%%%%%
%%%%%%%%%%%%%%%%%%%%%%%%%%%%%%%%%%%%%%%
\begin{figure*}[t]
\centering
\includegraphics[width=10cm]{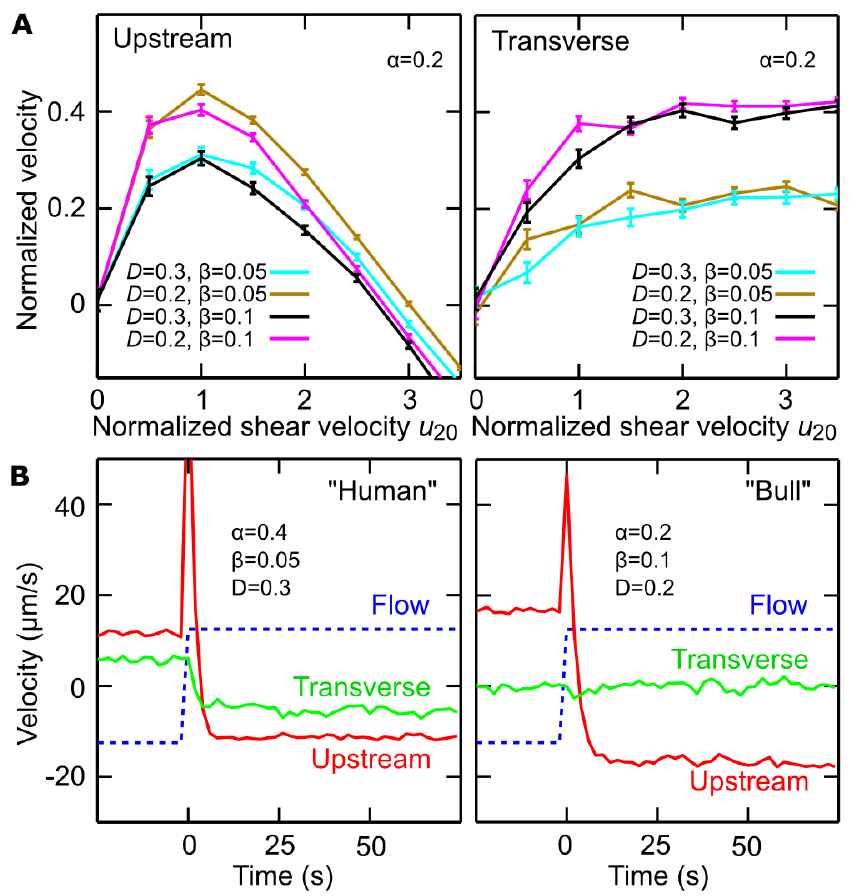}
\caption{
Model simulations reproduce main experimental observations.
(A) Upstream and transverse velocity for different values of the variability (effective noise) parameter $D$ in units 
rad$^2$/s and dimensionless shape factors $(\alpha,\beta)$. (B) Time response of a chiral swimmer with \mbox{$\chi=+1$} 
(\lq\lq Human\rq\rq) and a non-chiral swimmer with $\chi=0$ (\lq\lq Bull\rq\rq) to a reversal of the flow direction at 
time $t=0$. Blue dashed line shows fluid flow $u_y$ at $5\,\mu$m from the boundary. Simulation parameters ($N=1000$ 
trajectories, $A=10\,\mu$m, $\ell=60\,\mu$m, $V=50\,\mu$m/s) were chosen to match approximately those for viscosity 1$\,$mPa$\cdot$s 
in Fig. 2C. }
\end{figure*}
%%%%%%%%%%%%%%%%%%%%%%%%%%%%%%%%%%%%%%%
%%%%%%%%%%%%%%%%%%%%%%%%%%%%%%%%%%%%%%%
\newpage

\begin{figure*}[t]
\centering
\includegraphics[width=6.7cm]{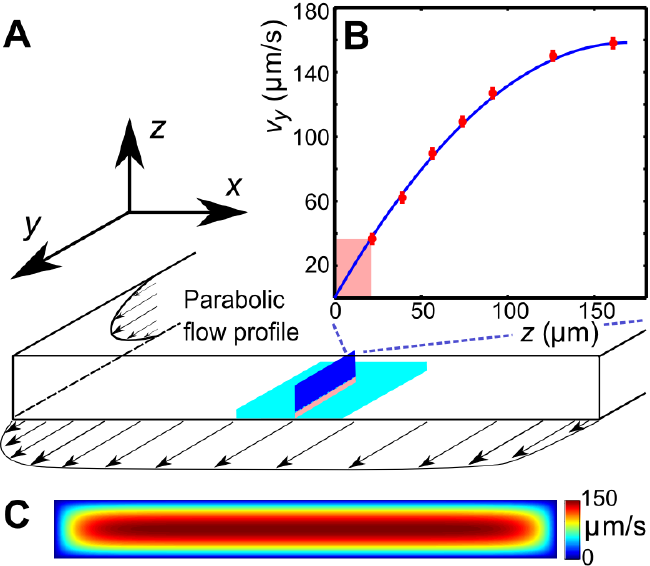}
\caption{ (A) Schematic of the microfluidic channel and field of view (turquoise region) in the sperm motility measurements. (B) Velocity profile at the center of the channel. Red symbols are values of the vertical velocity profile $u_y(z)$ measured by
PTV for the flow rate $0.1\mu$L/s. The solid line shows the theoretically calculated flow profile for the same flow 
rate. In motility experiments, values for the velocity gradient near the boundary (pink region) were obtained by measuring the flow velocity at $20$ $\mu$m from the boundary. (C)~Theoretical 2D flow speed profile in $(x,z)$-plane at flow rate $0.1\mu$L/s.
\label{fig:flow}
}
\end{figure*}

\end{document}